\documentclass[aps,prl,twocolumn,showpacs,superscriptaddress,longbibliography]{revtex4-1}
\usepackage{bm}
\usepackage{color} 
\usepackage{graphicx}
\usepackage{epsfig}
\usepackage{hyperref}
\usepackage{natbib}
\usepackage{amsmath}
\usepackage{amssymb}

\newcommand{\rmi}{{\rm i}}

\newcommand {\e}{{\rm e}}

\newcommand{\av}[1]{\left\langle #1\right\rangle}

\begin{document}

\title{Spatiotemporal spin fluctuations caused by spin-orbit-coupled Brownian motion}

\author{A.\,V.\,Poshakinskiy}\email{poshakinskiy@mail.ioffe.ru}
\affiliation{Ioffe Institute, St.~Petersburg 194021, Russia}
\author{S.\,A.\,Tarasenko}
\affiliation{Ioffe Institute, St.~Petersburg 194021, Russia}
\affiliation{St. Petersburg State Polytechnic University, St. Petersburg 195251, Russia}


\begin{abstract}
We develop a theory of thermal fluctuations of spin density emerging in a two-dimensional electron gas.
The spin fluctuations probed at spatially separated spots of the sample are correlated due to  
Brownian motion of electrons and spin-obit coupling. We calculate the spatiotemporal correlation functions of the spin density 
for both ballistic and diffusive transport of electrons and analyze them for different types of spin-orbit interaction including the isotropic Rashba model and persistent spin helix regime. The measurement of spatial spin fluctuations provides direct access to the parameters of spin-orbit coupling and spin transport in conditions close to the thermal equilibrium.
\end{abstract}

\maketitle

Thermal and quantum fluctuations of observables are canonical examples of stochastic processes which are inherent to physical systems.
Since the discovery of the random motion of pollen grains suspended in water by R. Brown followed by the theoretical works
by A. Einstein~\cite{Einstein1905} and M. Smoluchowski~\cite{Smoluchowski1906}, the study of fluctuations has become central to statistical physics and kinetics. The advances in optical spectroscopy have triggered the research of fluctuations of electron spins in atomic systems~\cite{Aleksandrov1981,Crooker2004} and, more recently, in semiconductor structures including bulk materials~\cite{Oestreich2005,Crooker2009,Romer2010}, quantum wells~\cite{Muller2008,Li2013,Poltavtsev2014}, quantum wires~\cite{Glazov2011,Agnihotri2012,Pershin2013}, and quantum dots~\cite{Crooker2010,Li2012,Dahbashi2012,Zapasskii2013,Smirnov2014}. Owing to the fundamental connection between fluctuations and dissipation processes, such a spin noise spectroscopy is becoming a powerful tool for studying the spin dynamics in conditions close to thermal equilibrium and for determining the spin relaxation times, $g$-factors, parameters of exchange and hyperfine interactions, etc., see recent review papers~\cite{Hubner2014,Zapasskii2013review}. 

The study of spin noise in semiconductors has been focused so far on the evolution of spin fluctuations in time (or the spectral
density of fluctuations) ~\cite{Oestreich2005,Crooker2009,Romer2010,Muller2008,Li2013,Poltavtsev2014,Glazov2011,Agnihotri2012,Crooker2010,Li2012,Dahbashi2012,Zapasskii2013,Smirnov2014}. However, already in the first experiments on spin noise in an electron gas it was found that the dynamics of spin fluctuations is sensitive to the spatial size of the probed area of the sample due to diffusion of electrons out of this area~\cite{Muller2008}. Moreover, the general analysis shows that the temporal and spatial correlations of spin fluctuations emerging in an electron gas are coupled due to Brownian motion of electrons and spin-orbit interaction. Therefore, a natural and consistent description of the spin noise in an electron gas is spatiotemporal. Meanwhile, such a study of the spin noise has been limited up to now to one-dimensional systems~\cite{Pershin2013} in which the motion of an electron in the real space and its spin rotation are locked and the spin dynamics is rather obvious. In systems of higher dimensions, the spin dynamics is complicated by the diversity of electron trajectories and
the one-to-one correspondence between the real-space shift of electrons and the spin rotation angle vanishes. Here, we develop a theory of propagating spin fluctuations in a two-dimensional electron gas confined in a quantum well (QW).
We calculate the spatiotemporal correlation functions of the spin density for both diffusive and ballistic regimes and show that the correlation functions provide a direct information on the spin-orbit coupling in an electron gas and parameters of the spin transport. We show that the correlations of spin fluctuations drastically increase in the regime of persistent spin helix~\cite{Bernevig2006}, which suggests a noninvasive method for probing this intriguing phenomenon. Since the access to electron spin density with high spatial and temporal resolutions is experimentally available nowadays~\cite{Cameron1996,Weber2007,Volkl11}, the measurements of fluctuations will enable the study of spin waves and spin diffusion processes in conditions close to the thermal equilibrium.

\begin{figure}[b]
\includegraphics[width=0.7\columnwidth]{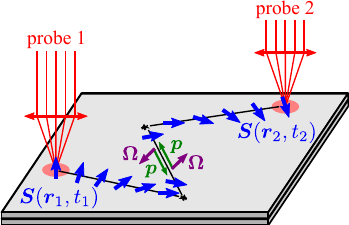}
\caption{Probe-probe measurements of the spatial correlations of spin fluctuations in an electron gas. Fluctuations of the spin density at the points $\bm{r}_1$ and $\bm{r}_2$ are correlated due to Brownian motion of electrons. The sign of the correlation function, positive or negative, depends on the distance $\bm r_1 - \bm r_2$ and the frequency $\bm \Omega (\bm p)$ of spin precession in 
the effective magnetic field which acts upon the electron spins when electrons walk.}
\label{sketch}
\end{figure}

The Gedankenexperiment we consider is illustrated in Fig.~\ref{sketch}. A two-dimensional electron gas confined in a semiconductor quantum well is in thermal equilibrium. The gas is spin unpolarized, however there are incessant fluctuations of the  spin density. The spin fluctuations propagate in the QW plane due to Brownian motion of electrons and precess in the effective magnetic field caused by spin-orbit coupling. In quantum wells, the frequency of spin precession $\bm \Omega$ depends linearly on the electron momentum $\bm p$ 
which leads to a relationship between the electron trajectory in real space and spin rotation angle. As a result, the Brownian motion of electrons leads to correlations of the spin density probed at different spots of the sample. Whether the spin fluctuations at the points $\bm r_1$ and $\bm r_2$ are positively or negatively correlated depends on the average spin rotation angle when electrons walk between $\bm r_1$ and $\bm r_2$ points which depends on the distance $\bm r = \bm r_1 - \bm r_2$ and the effective magnetic field strength. 

To characterize the fluctuations we introduce the correlation functions of the spin density $\bm S(\bm r_1,t_1)$ and $\bm S(\bm r_2,t_2)$ at different points of the sample. For spatially and temporally homogenous systems, the correlation functions depend on the time difference and coordinate difference and are defined by
\begin{equation}\label{def}
  K_{\alpha\beta}(\bm r_1 - \bm r_2, t_1 - t_2) = \av{ \left\{ \hat S_\alpha(\bm r_1,t_1) , \hat S_\beta(\bm r_2,t_2) \right\} } \:,
\end{equation}
where $\hat S_\alpha(\bm r_1,t_1) = \hat\psi^\dag(\bm r_1,t_1)\sigma_\alpha \hat\psi(\bm r_1, t_1)$ is the spin density operator, $\hat \psi(\bm r_1,t_1)$ is the electron-field operator, $\sigma_\alpha$ are the Pauli matrices, $\{A,B\}=(AB+BA)/2$ is the symmetrized product of the operators, and the angular brackets denote averaging with the quantum-mechanical density matrix corresponding to thermodynamic equilibrium. The necessity to use the symmetrized product of $\hat S_\alpha(\bm r_1,t_1)$ and $\hat S_\beta(\bm r_2,t_2)$ is caused by the fact that different components of the spin density operator do not commute with each other. The correlation functions
satisfy the relations $ K_{\alpha\beta}(\bm r, t) =  K_{\alpha\beta}^*(\bm r, t)$, $ K_{\alpha\beta}(\bm r, t) =  K_{\beta\alpha}(-\bm r, -t)$, and $ K_{\alpha\beta}(\bm r, t) =  K_{\alpha\beta}(\bm r, -t)$; the latter is due to time inversion symmetry in the absence of an external magnetic field.

We calculate the correlation functions by using the fluctuation-dissipation theorem~\cite{LandauLifshits5} that relates the  
Fourier components of the correlation functions $K_{\alpha\beta}(\bm q, \omega) = \int \int K_{\alpha\beta}(\bm r, t) \, \e^{\rmi\omega t - \rmi\bm q \cdot \bm r} dt\, d\bm r $ to the Fourier components of the spin susceptibility $\chi_{\alpha\beta}(\bm q, \omega)$,
\begin{equation}\label{FDT}
  K_{\alpha\beta}(\bm q, \omega) =  \frac{\chi_{\alpha\beta}(\bm q, \omega)-\chi^*_{\beta\alpha}(\bm q, \omega)}{2\rmi\hbar} \coth \frac{\hbar \omega}{2 T} \,.
\end{equation}
Here, $\chi_{\alpha\beta}(\bm q, \omega)$ is defined as the linear response, $S_\alpha(\bm q, \omega) = \chi_{\alpha\beta}(\bm q, \omega) F_\beta(\bm q, \omega)$,  of the spin density $S_\alpha(\bm q, \omega)$ to the ``force'' $F_\beta(\bm q, \omega)$ whose action upon the system is described by the Hamiltonian perturbation $\hat V = -\int \, \hat{\bm S}(\bm r,t)\cdot \bm F(\bm r, t) d\bm r$, and $T$ is the temperature.

The susceptibility $\chi_{\alpha\beta}(\bm q, \omega)$ is readily expressed via retarded and advanced electron Green's functions $G^{R,A}_{\varepsilon}(\bm r,\bm r')$ as follows
\begin{align}\label{GR_GA}
  &\chi_{\alpha\beta}(\bm q, \omega) =  \frac{m^*}{4\pi\hbar^2} \delta_{\alpha\beta} - \frac{\rmi}{4} \int\limits_{0}^{\hbar\omega}\frac{d\varepsilon}{2\pi} \int d\bm r \int d\bm r' \\
  &\times {\rm Tr}\av{\sigma_\alpha G^R_{\varepsilon}(\bm r, \bm r')\sigma_\beta G^A_{\varepsilon-\hbar\omega}(\bm r',\bm r)} \e^{\rmi \bm q (\bm r'-\bm r)}, \nonumber
\end{align}
where $m^*$ is the effective mass, $\delta_{\alpha\beta}$ is the Kronecker delta, the angular brackets denote averaging over the disorder, and we assume that the electron gas is degenerate, $\hbar q \ll p_F$ and $\hbar\omega \ll \varepsilon_F$, with $p_F$ and $\varepsilon_F$ being the Fermi momentum and the Fermi energy, respectively.  

Averaging over the positions of impurities leads to the sum of the ladder diagrams. For the case of short-range scattering, the sum of the ladder diagrams has the form
\begin{align}\label{laddersum}
 \chi_{\alpha\beta}(\bm q, \omega) =  \frac{m^* \delta_{\alpha\beta}}{4\pi\hbar^2} - \frac{\rmi}{4} \int\limits_{0}^{\hbar\omega}\frac{d\varepsilon}{2\pi} {\rm Tr} \left[ \sum_{\bm p}  \sigma_\alpha G^R_{\bm{p},\varepsilon} \sigma_\beta G^A_{\bm{p}-\hbar\bm{q},\varepsilon-\hbar\omega} \right. 	\nonumber \\
  +\frac{\hbar^3}{m^* \tau} \left. \sum_{\bm p \bm{p}'}  \sigma_\alpha G^R_{\bm{p},\varepsilon} G^R_{\bm{p}',\varepsilon} \sigma_\beta G^A_{\bm{p}'-\hbar\bm{q},\varepsilon-\hbar\omega} G^A_{\bm{p}-\hbar\bm{q},\varepsilon-\hbar\omega} + \ldots \right] ,
\end{align}
where $G^{R,A}_{\bm p, \varepsilon}$ are impurity-averaged Green's functions,
\begin{equation}
G^{R,A}_{\bm p, \varepsilon}=\frac{1}{ \varepsilon_F + \varepsilon - {\bm p}^2/2m^*  - (\hbar/2)\bm\Omega(\bm p)\cdot \bm\sigma  \pm \rmi \hbar/2\tau} \,,
\end{equation}
$\tau$ is the relaxation time, $\bm\Omega(\bm p)$ is the Larmor frequency corresponding to the effective magnetic field.

Straightforward calculation of the series~\eqref{laddersum} yields
\begin{equation}\label{chi}
\chi_{\alpha\beta}(\bm q, \omega) = \frac{m^*}{4\pi\hbar^2} \left[\delta_{\alpha\beta}+\rmi\omega T_{\alpha\beta}(\bm q, \omega)\right] \,,
\end{equation}
where $\bm T(\bm q, \omega)$ is the spin lifetime tensor,
\begin{equation}\label{T}
\bm T(\bm q, \omega) =   \tau \bm C(\bm q, \omega) [1-\bm C(\bm q, \omega)]^{-1} \,,
\end{equation}
$\bm C(\bm q, \omega)$ is the matrix  is given by 
\begin{equation}\label{A}
  C_{\alpha\beta}(\bm q,\omega) =  \int \frac{d \varphi_{\bm p}}{2\pi}\, \frac{\eta\delta_{\alpha\beta}-\epsilon_{\alpha\beta\gamma} \Omega_{\gamma}\tau+\Omega_{\alpha} \Omega_{\beta}\tau^2/\eta}{\eta^2+\Omega^2 \tau^2} \,,
\end{equation}
$\eta(\bm p) =1-{\rm i}\omega\tau + {\rm i} (\tau/m^*) \bm q \cdot \bm p$, $\epsilon_{\alpha\beta\gamma}$ is the Levi-Civita tensor,
$\varphi_{\bm p}$ is the polar angle of the momentum $\bm p$, and $|\bm p|=p_F$. We note that the tensor $\bm T(\bm q,\omega)$ describes also the spin density $\bm S(\bm q,\omega)$ emerging in the sample when the spin is generated at the Fermi level at the rate $\bm G(\bm q,\omega)$, $S_{\alpha} = T_{\alpha\beta} G_\beta$. Calculation of the spin lifetime tensor $\bm T$ in the framework of the Boltzmann
equation in stationary and spatially homogenous case for QWs of different symmetry and arbitrary $\Omega \tau$ is described in Refs.~\cite{Poshakinskiy11,Poshakinskiy2013}.

Combining Eqs.~\eqref{FDT} and~\eqref{chi} we obtain the correlation functions at $T \gg \hbar\omega$
\begin{equation}\label{K}
K_{\alpha\beta}(\bm q, \omega)  =  \frac{m^* T}{4\pi\hbar^2}   [  T_{\alpha\beta}(\bm q, \omega) +   T_{\beta\alpha}^*(\bm q, \omega) ]\,.
\end{equation}
Equation~\eqref{K} describes the spatial and temporal correlations of spin density fluctuations emerging in two-dimensional electron gas for arbitrary form of the effective magnetic field and arbitrary parameter $\Omega\tau$. Below, we discuss the correlations for the ballistic ($\Omega \tau \gg 1$) and diffusive ($\Omega \tau \ll 1$) regimes of spin dynamics. We consider (001)-oriented QWs where the $\bm p$-linear spin-orbit coupling is described by the effective magnetic field
\begin{equation}
\bm\Omega(\bm p) = [(\Omega_D+\Omega_R) p_y/p_F, (\Omega_D-\Omega_R)p_x/p_F,0] \:,
\end{equation}
$\Omega_D$ and $\Omega_R$ are the Dresselhaus and Rashba field strengths at the Fermi level, respectively, $x\parallel [1\bar 10]$ and $y\parallel [110]$ are the crystal-structure-enforced eigen axes in the QW plane, and $z$ is the QW normal~\cite{Ivchenko_book}.

\begin{figure}[t]
\includegraphics[width=1\columnwidth]{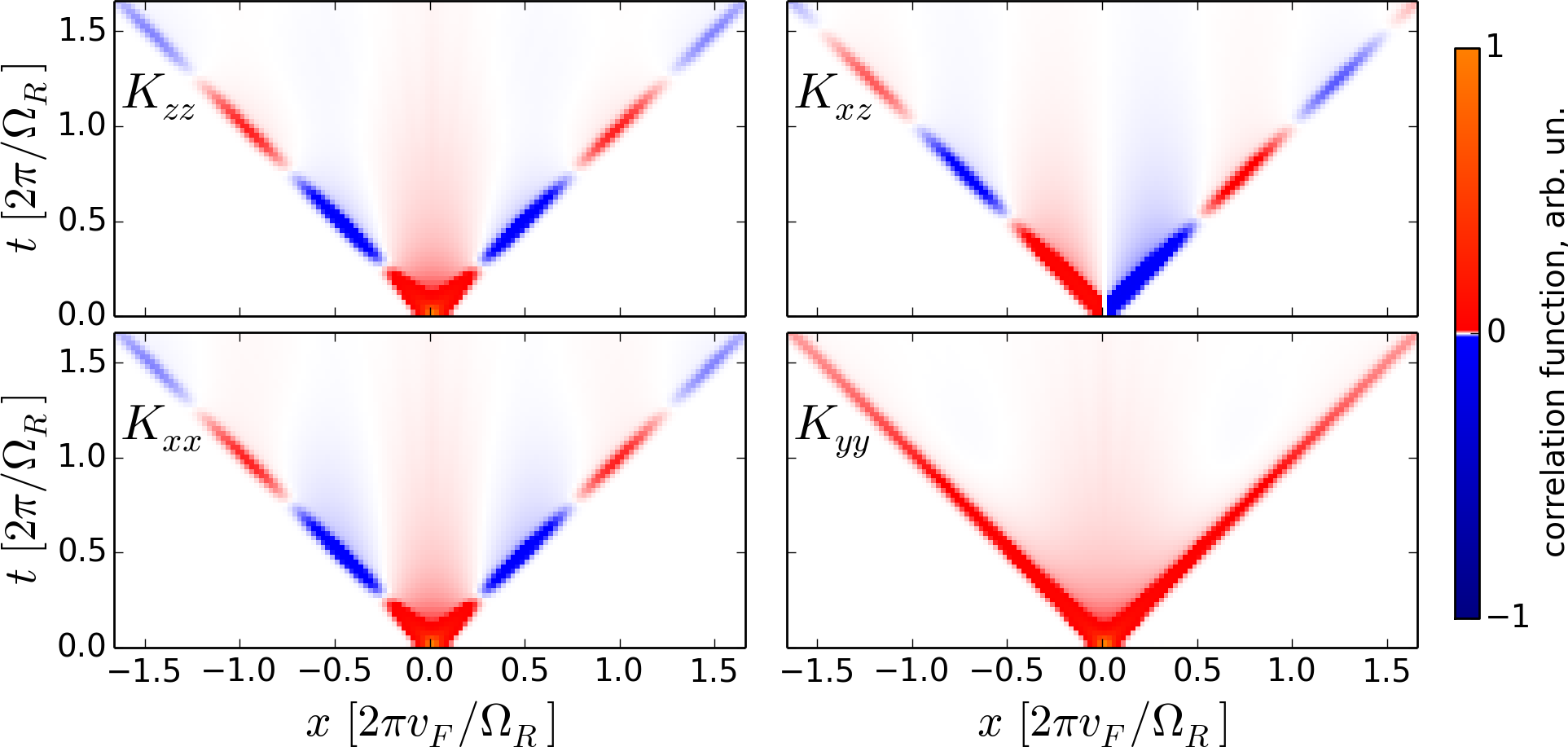}
\caption{Spin density correlation functions $K_{\alpha\beta}(\bm r,t)$ for the Rashba spin-orbit interaction and coordinate difference $\bm r \parallel x$ between the points where the spin fluctuations are probed. Map are calculated for $\Omega_R\tau =5$
which is close to ballistic transport of electrons between the points.\vspace{-.2cm}}
\label{figure2}
\end{figure}

Figure~\ref{figure2} shows the coordinate and time dependence of the correlation functions $K_{\alpha\beta} (\bm r, t)$ for  ballistic
transport of electrons between the spots where the spin fluctuations are probed. The maps are calculated for the Rashba effective magnetic field. The pronounced correlations of the spin fluctuations probed at the spots separated by the distance $\bm r$ emerge at the time delay $t=(m^*/p_F)  r$ that follows from the ballistic behavior of electron transport. The correlation function contains both the diagonal $K_{\alpha\alpha}$ and off-diagonal $K_{\alpha\beta}$ ($\alpha\neq\beta$) components. The latter originate from the precession of electron spins in the effective magnetic field when electrons propagate between the spots. The precession is also responsible for the oscillatory behavior of the correlation functions. Figure~\ref{figure2} is plotted for $\bm r \parallel x$. Accordingly, the Rashba field for electrons propagating between the spots points along the $y$ axis and couples the $x$ and $z$ components of the spin density. The frequency of spin precession is given by $\Omega_R$ which results in the oscillations of the correlation functions in real space with the wave vector $q = \Omega_R m^*/p_F$.

In the case of arbitrary effective field $\bm{\Omega}(\bm p)$, the correlation function in the ballistic limit has the form  
\begin{equation}\label{ballistic}
\bm K(\bm r, t)  = \frac{m^* T}{8 \pi^2 \hbar^2 \,r}   \bm R\left[\bm{\Omega}\left(\frac{m^* \bm r}{t}\right) t\right]\, \delta(r - v_F |t|)  \,,
\end{equation}
where $\bm R[\bm \Theta]$ is the matrix of rotation by the angle $\bm \Theta$ and $v_F = p_F/m^*$ is the Fermi velocity. In particular, the $K_{zz}$ component  for (001)-oriented QWs assumes the form
\begin{equation}\label{ballistic001}
\bm K_{zz}  =  \frac{m^* T}{8 \pi^2 \hbar^2 \,r} \cos \left( \Omega_{\Sigma} t \sqrt{1 - \sin 2\phi \cos 2\varphi_{\bm r}} \right)  \delta (r - v_F |t|)  \,,
\end{equation}
where $\Omega_{\Sigma}=\sqrt{\Omega_D^2 + \Omega_R^2}$, $\phi = \arctan (\Omega_R/\Omega_D)$, and $\varphi_{\bm r} =  \arctan (y/x)$ is the polar angle of the vector $\bm r$.

\begin{figure}[t]
\includegraphics[width=0.99\columnwidth]{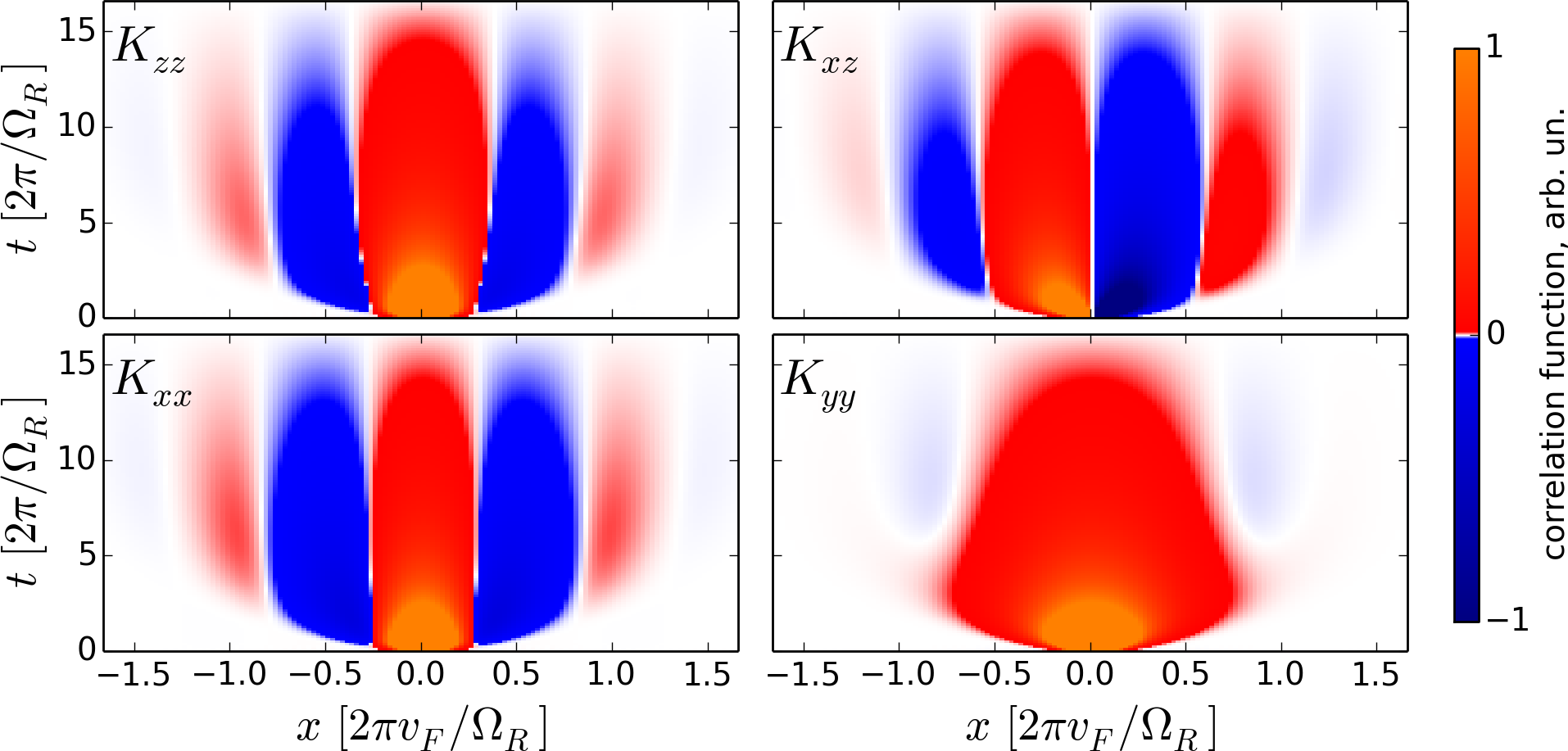}
\caption{Spin density correlation functions $K_{\alpha\beta}(\bm r,t)$ for the Rashba spin-orbit interaction and coordinate difference $\bm r \parallel x$ between the points where the spin fluctuations are probed. Maps are calculated for $\Omega_R\tau =0.2$ which corresponds to diffusive transport of electrons between the points.\vspace{-.2cm}}
\label{figure3}
\end{figure}

Now we turn to diffusive transport of electrons. The calculated dependence of the correlation functions $K_{\alpha\beta} (\bm r,t)$ on coordinate and time for this regime is shown in Fig.~\ref{figure3}. The fluctuations of spin density at the spots separated by the distance $\bm r$ get correlated at the time delay $t \gtrsim r^2/D$, where $D = v_F^2 \tau /2$ is the diffusion coefficient. Despite the diffusive transport, for which electrons can travel along many different trajectories, the correlation functions do contain oscillations as a function of distance. This can be attributed to the fact that the major contribution to correlations is given by the diffusive trajectories which are close to the straight line connecting the spots.

An analytical equation for the correlation function in the diffusive regime can be derived from Eqs.~\eqref{T}-\eqref{K} by considering the case $\Omega \tau \ll 1$ and, hence, $\omega\tau, Dq^2\tau  \ll 1$. Such a procedure  yields
\begin{equation}\label{Tcoldom}
 \bm K(\bm q,\omega) = \frac{m^* T}{4\pi\hbar^2}\left( \frac{1}{ - \rmi\omega + \bm\Gamma + D q^2  +  2\rmi D \bm \Lambda(\bm q)} + {\rm H.c.} \right),
\end{equation}
where $\Gamma_{\alpha\beta} = \tau\int  (\Omega^2-\Omega_\alpha\Omega_\beta )\,d\varphi_{\bm p} /(2\pi ) $ is the  D'yakonov--Perel' spin-relaxation-rate tensor for homogeneous spin distribution~\cite{Dyakonov86}  and $\Lambda_{\alpha\beta}(\bm q) = \int \epsilon_{\alpha\gamma\beta} \Omega_\gamma\, \bm q \cdot \bm p\, d\varphi_{\bm p} /(2\pi m^*)$. The tensor $\Lambda_{\alpha\beta}(\bm q)$ describes the precession of electron spin at diffusion. At large delay times, the correlations are determined by the spin excitations with the longest lifetime. For pure Rashba or Dresselhaus spin-orbit coupling, the $K_{zz}$ component of the correlation function at $t \gg 1/\Omega^2\tau$ assumes the form
\begin{equation}
K_{zz}(\bm r, t) = \frac{3m^* \Omega T}{32\pi v_F\hbar^2} \frac{\e^{-(7/32)\Omega^2\tau |t|}}{\sqrt{4\pi D |t|}} J_0\left( \sqrt\frac{15}{16} \frac{\Omega r}{v_F}\right) ,
\end{equation}
where $J_0$ is the Bessel function.
The emergence of long-lived spin polarization waves with the wave vector $q = \sqrt{15/16}\, \Omega /v_F$ 
under inhomogeneous spin pumping was theoretically predicted by V.A.~Froltsov~\cite{Froltsov2001} and observed by means of transient spin-grating spectroscopy by C.P.~Weber {\it et al.}~\cite{Weber2007}. 

\begin{figure}[t]
\includegraphics[width=0.99\columnwidth]{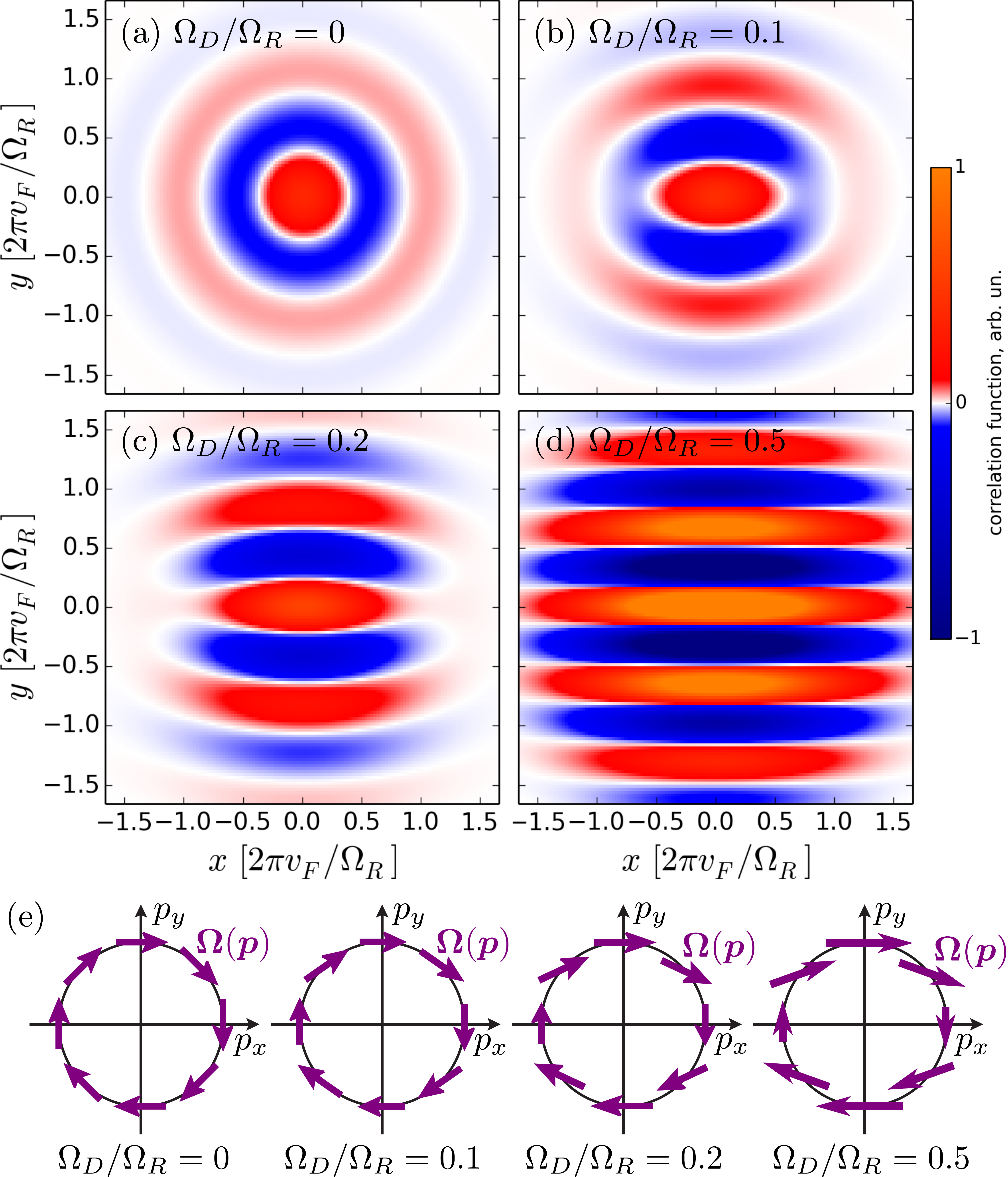}
\caption{(a)-(d) Correlation function of the out-of-plane spin fluctuations $K_{zz}$ in real space in (001)-grown QW. Maps are calculated for the delay time $t = 20 /(\Omega_R^2 \tau)$, $\Omega_R\tau=0.2$, and different ratio of the Dresselhaus and Rashba terms. (e) Distributions of the effective magnetic field on the Fermi circle for different ratio of the Dresselhaus and Rashba terms.}
\label{figure4}
\end{figure}

Finally, we discuss the spin fluctuations in the presence of spin splitting anisotropy, caused by interference of the Rashba and Dresselhaus terms, which can be very pronounced in asymmetric (001)-oriented QWs~\cite{Ganichev2014}.
Figure~\ref{figure4} shows the dependence of the correlation function $K_{zz}$ on the relative position of the probed spots in the QW plane. The maps are calculated for different ratio of the Dresselhaus and Rashba fields. The anisotropy of spin splitting in the momentum space leads to an anisotropy of the correlation function in real space which turns out to be strong even for small ratio $\Omega_D/\Omega_R$, cf. Figs.~\ref{figure4}a-d and~\ref{figure4}e. Another striking consequence of the interplay of the Rashba and Dresselhaus fields is a drastic increase in the lifetime and amplitude of spin correlations, see Fig.~\ref{figure4}. This is due to the fact that, at equal strengths of the Rashba and Dresselhaus fields, the SU(2) spin rotation symmetry emerges in the system leading to an appearance of spin density waves with infinite lifetime (persistent spin helix)~\cite{Bernevig2006,Koralek2009,Walser2012,Sasaki2014,Ishihara2014}. We note that the emergence of long-lived waves with the wave vector along the $y$ axis can be clearly seen already at $\Omega_D/\Omega_R=0.5$, Fig.~\ref{figure4}d.
An advantage of the spin noise spectroscopy for the study of spin helix as compared to the pump-probe technique is that no photoexcited carriers and, hence, no additional mechanisms of spin dephasing are introduced.

At large delay times, the correlation function is determined by the spin density waves with the longest lifetime. Such waves are directed along the $x$ or $y$ axis depending on the sign of the product $\Omega_R \Omega_D$. At $\Omega_R \Omega_D >0$, the correlation function $K_{zz}$ oscillates along the $y$ axis and Eq.~\eqref{Tcoldom} yields its asymptotic behavior   
\begin{equation}
K_{zz} (\bm r,t) \propto T \, \frac{\e^{-\tilde{\gamma} t} \cos (\tilde{q} y)}{t} \:,
\end{equation}
where the wave vector $\tilde q$ and the decay rate $\tilde\gamma$ are given by
\begin{equation}
 \tilde q = \frac{\Omega_R+ \Omega_D}{v_F} \sqrt{ 1 - \frac{1}{16} \tan^4 \left(\phi-\frac{\pi}{4}\right)}\:,
\end{equation}
\[
\tilde\gamma = \frac{9(\Omega_R-\Omega_D)^2 \tau}{32} \left( 1 - \frac{2}{9} \frac{1}{1+\sin 2\phi} \right) \:.
\]
While the decay rate of the correlation function is mostly determined by the difference of the strengths of the Rashba and Dresselhau fields, the period of oscillations in real space is determined by the sum of the field strengths.
 
To summarize, we have developed a theory of spatiotemporal fluctuations of spin density emerging in a two-dimensional
electron gas with spin-orbit coupling. We have calculated the correlation functions of spin density for both ballistic and diffusive regimes of electron transport and analyzed them for different types of spin-orbit coupling that can be realized in quantum wells.
The correlations of spin fluctuations at large delay times are determined by the long-lived waves of spin density and drastically increas in the regime of persistent spin helix. 

Financial support by the RFBR, RF President Grants No. MD-3098.2014.2 and No. NSh-1085.2014.2, EU project SPANGL4Q, and the ``Dynasty'' Foundation is gratefully acknowledged.

\end{document}